\documentclass[11pt]{article}

\usepackage[authoryear,round]{natbib}
\usepackage{amsmath,amssymb,amsfonts,amsthm,bbm,mathtools}
\usepackage{epic,eepic,epsfig,longtable}
\usepackage{multirow,verbatim}
\usepackage{array}
\usepackage{kotex}
\usepackage{todonotes}
\usepackage{epsfig}
\usepackage{setspace}
\usepackage{CJK}
\usepackage{color}
\usepackage[hyphens]{url}

\usepackage{rotating}
\usepackage{enumitem}

\usepackage[unicode=true]{hyperref}
\hypersetup{breaklinks=true,
            colorlinks=false,
            pdfborder={0 0 0}}

\makeatletter
\def\singlespace{\def\baselinestretch{1}\@normalsize}

\makeatletter
\def\singlespace{\def\baselinestretch{1}\@normalsize}

\numberwithin{equation}{section}

\newcommand{\bfm}[1]{\ensuremath{\mathbf{#1}}}

   \def\bD{\bfm D}

\def\calN{{\cal  N}}

\def\calS{{\cal  S}}

\newcommand{\bfsym}[1]{\ensuremath{\boldsymbol{#1}}}

 \def\bfeta{\bfsym {\eta}}

\def\today{\ifcase\month\or
  January\or February\or March\or April\or May\or June\or
  July\or August\or September\or October\or November\or December\fi
  \space\number\day, \number\year}

\newdimen\biblioindent    \biblioindent=30pt

 at 8truept

\newcommand{\beq}{\begin{equation}}
  \newcommand{\eeq}{\end{equation}}
\newcommand{\beqn}{\begin{eqnarray}}
  \newcommand{\eeqn}{\end{eqnarray}}
\newcommand{\beqnn}{\begin{eqnarray*}}
  \newcommand{\eeqnn}{\end{eqnarray*}}

\long\def\comment#1{}

\allowdisplaybreaks
\usepackage{verbatim}
\renewcommand{\baselinestretch}{1.66}
\newtheorem{thm}{Theorem}

\newcounter{CondCounter}

\theoremstyle{definition}

\newtheorem{remark}{Remark}

\def \etadot   {\eta_{\boldsymbol{\cdot}}}

\begin{document}

\title{$L_1$ Prominence Measures for Directed Graphs}

\author{
  Seungwoo Kang\footnote{kangsw0401@snu.ac.kr}~ and Hee-Seok Oh\footnote{Corresponding author: heeseok@stats.snu.ac.kr}\vspace{0.1in}\\
  Department of Statistics, Seoul National University\\
  Seoul 08826, Republic of Korea}
\date{\today}

\maketitle

\begin{abstract}
\noindent
We introduce novel measures, $L_1$ prestige and $L_1$ centrality, for quantifying the prominence of each vertex in a strongly connected and directed graph by utilizing the concept of $L_1$ data depth (Vardi and Zhang, Proc.\ Natl.\ Acad.\ Sci.\ U.S.A.\ 97(4):1423--1426, 2000). The former measure quantifies the degree of prominence of each vertex in receiving choices, whereas the latter measure evaluates the degree of importance in giving choices. The proposed measures can handle graphs with both edge and vertex weights, as well as undirected graphs. However, examining a graph using a measure defined over a single `scale' inevitably leads to a loss of information, as each vertex may exhibit distinct structural characteristics at different levels of locality. To this end, we further develop local versions of the proposed measures with a tunable locality parameter. Using these tools, we present a multiscale network analysis framework that provides much richer structural information about each vertex than a single-scale inspection. By applying the proposed measures to the networks constructed from the Seoul Mobility Flow Data, it is demonstrated that these measures accurately depict and uncover the inherent characteristics of individual city regions.
\end{abstract}

\section{Introduction}

With recent advances in science and technology, modern data often take the form of a network representing individual units and their relationships through nodes (also known as vertices or points) and edges (also known as lines, links, or connections) in a graph. Some examples include social networks \citep{wasserman_faust_1994}, the World Wide Web \citep{page1998pagerank,kleinberg1999authoritative}, and biological networks such as protein networks and food webs \citep{girvan2002community,jonsson2006cluster}. In network studies, there is a fundamental interest in quantifying the prominence of each vertex based on its involvement in the network. For undirected graphs, a popular tool is the \emph{centrality} measure. For example, \emph{degree centrality} is a measure that counts the number of edges connected to each vertex; a vertex connected to a large number of edges has a high degree centrality, indicating its importance. There are many different centrality measures, each of which has a solid historical basis in social network analysis \citep{freeman1978centrality}. See, for example, \cite{knoke1983}, \citet[Ch.5]{wasserman_faust_1994}, and \cite{borgatti2006graph} for reviews.

When analyzing directed graphs, the prominence of each node can be evaluated from two distinct perspectives: its importance in \emph{receiving} choices and its importance in \emph{giving} choices. The former notion is referred to as \emph{prestige}, whereas the latter is known as \emph{centrality} \citep[Ch.5]{wasserman_faust_1994}. For instance, we can extend degree centrality to directed graphs by counting the number of incoming edges (in-degree) and outgoing edges (out-degree) for each vertex. Measures in \citet[Ch.9]{nooy2011exploratory}, \cite{katz1953new}, \cite{hubbell1965input}, \cite{bonacich1987power}, \cite{page1998pagerank}, and \cite{kleinberg1999authoritative} are well-known examples. Alternatively, some measures \citep[e.g.,][]{white1994betweenness} quantify the prominence of each node in a directed graph according to its role as a broker or a mediator. However, this viewpoint is not of interest in this paper. 

From a statistical point of view, computing a prominence measure (such as centrality or prestige) can be regarded as reducing the dimensionality of a large graph dataset to a small set of numbers because each vertex is assigned a single value. Consequently, there is an inevitable loss of information when evaluating prominence. For example, the importance of a vertex may vary depending on whether the entire graph's vertices or just relevant ones---such as those close to the vertex of interest---are taken into consideration. This motivates \emph{multiscale analysis}, whose core philosophy is well described in the scale-space theory of computer vision \citep{lindeberg1994scale,chaudhuri2000scale}. As \cite{lindeberg1994scale} wrote, 
\begin{quote}
``An inherent property of objects in the world is that they only exist as meaningful entities over certain ranges of scale. If one aims to describe the structure of unknown real-world signals, then a multi-scale representation of data is of crucial importance.''
\end{quote}
Applying this perspective to the concept of prominence measures, examining the network at different `scales' can extract much richer structural information. Indeed, this idea is shared by a few earlier works. A method proposed by \cite{kang-l1cent-paper} identifies the neighborhood of a vertex based on its prominence measure and evaluates the local prominence among its neighbors; however, this approach is limited to undirected graphs. We also note that works by \cite{bonacich1987power}, \cite{agneessens2017geodesic}, and \cite{everett2022extended} utilize a tuning parameter to control the relative influence of nearby nodes over distant ones in their prominence measures. Yet, the interpretation of the tuning parameters in these studies remains ambiguous.

In this study, we propose two new prominence measures, termed \emph{$L_1$ prestige} and \emph{$L_1$ centrality}, which assess the `global' prominence of each node when receiving and giving choices in strongly connected directed graphs. Section \ref{sec:l1pres} provides the definitions and properties of the measures. 
Moreover, localized versions of the two proposed measures are developed in Section \ref{sec:multiscale}, with a locality parameter that can be varied and easily interpreted. 
Our proposed global and local measures have several advantages compared to existing centrality and prestige measures.
\begin{itemize}
\item[(a)] The proposed measures can handle graphs with vertex and edge weights. Vertex weights can be regarded as contextual information about nodes' importance \citep{schoenfeld2021shortest} or prior knowledge about the social prestige of each node \citep[Ch.9]{nooy2011exploratory}. Moreover, the proposed measures can take into account edge weights, which encode the strength of the relationship between nodes. Some measures in the literature cannot handle vertex or edge weights. 

\item[(b)] Our measures enable a multiscale analysis of a single network, which can reveal structural information that cannot be detected by a single prominence measure. To the best of our knowledge, this is the first approach to formalize a multiscale analysis framework for prestige and centrality measures in directed graphs. 

\item[(c)] The proposed measures are canonical extensions of the $L_1$ centrality and the local $L_1$ centrality for undirected graphs in \cite{kang-l1cent-paper}. Hence, the proposed measures form a family of prominence measures applicable to both directed and undirected graphs.
\end{itemize}


To illustrate our method, we consider networks built from `Seoul Mobility Flow Data.' This is an example of an urban flow network encompassing urban taxi trajectories \citep{huang2015trajgraph} and bike-sharing networks \citep{wu2020analyzing}. The Seoul Mobility Flow Data contains information about the number of individuals moving in Seoul, categorized by various demographic criteria. The raw dataset is massive, owing to the enormous number of movements in Seoul. For instance, the dataset for December 2023 contains more than 134 million records. 

\begin{figure}
\center
\includegraphics[width=.49\textwidth]{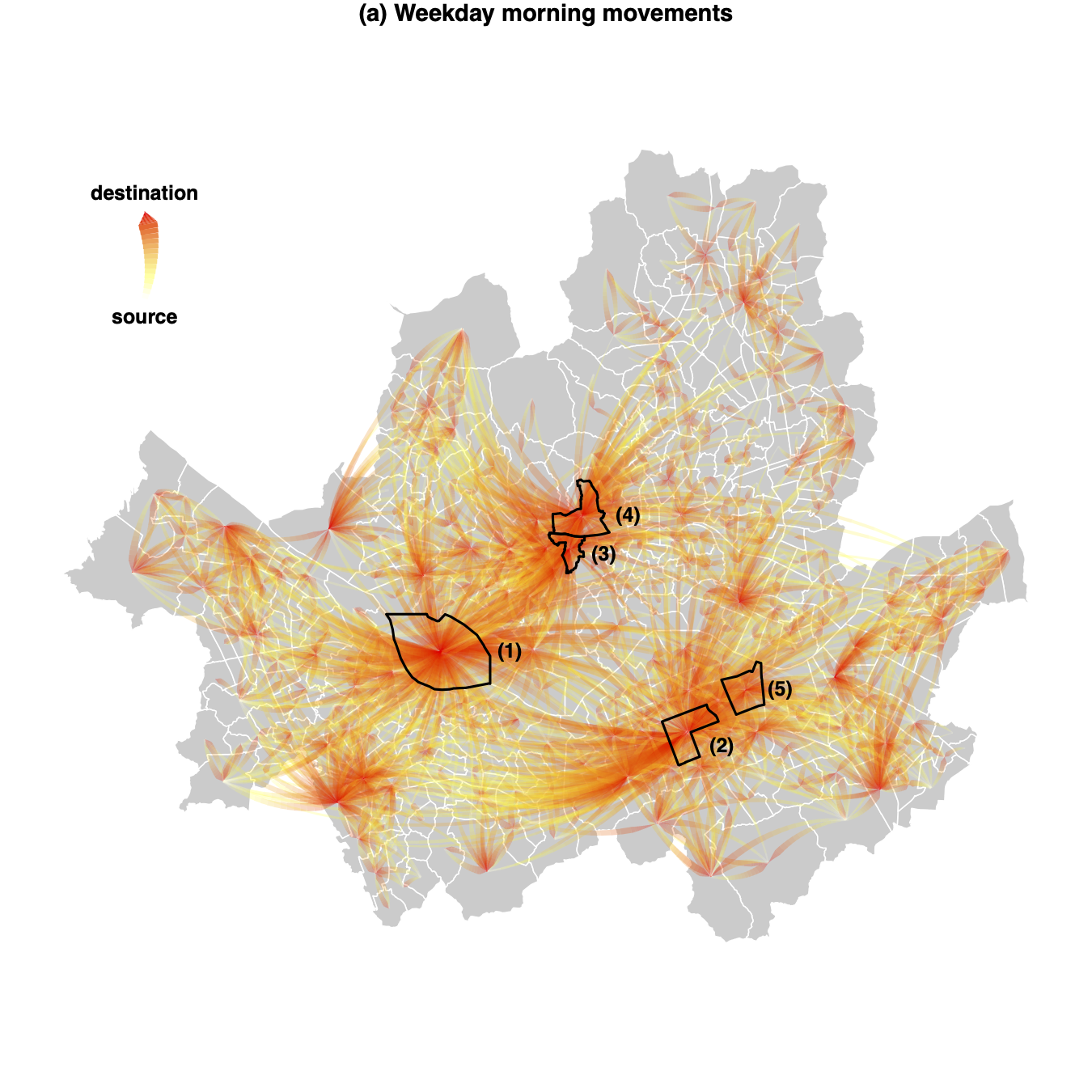}
\includegraphics[width=.49\textwidth]{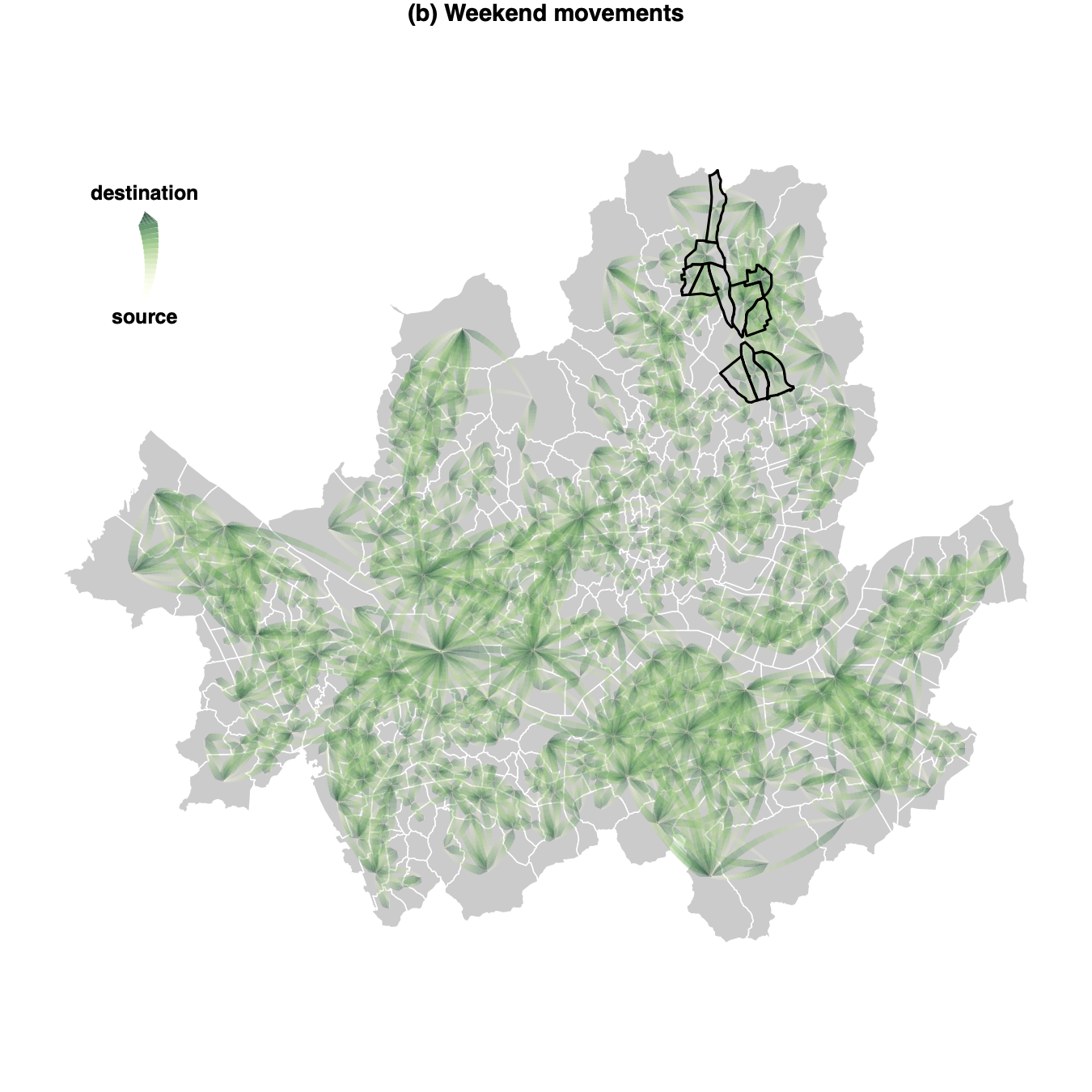}\vspace{-10mm}
\caption{(a) Visualization of weekday morning movements in Seoul during December 2023. An arrow is drawn if the number of individuals moving from one town to another exceeds 2000 (people). Movements below 2000 are omitted for clear visualization. The proposed $L_1$ prestige measure can be applied to the directed graph constructed from this dataset, where each vertex corresponds to one of the 424 towns in Seoul. The five regions with the highest $L_1$ prestige are shown on the plot by black border lines, and the numbers indicate the order (e.g., (2) is the region with the second largest $L_1$ prestige). (b) Visualization of weekend movements above 4000 in Seoul during December 2023. By conducting the multiscale network analysis of the directed graph constructed from this dataset, one can detect regions that show distinct structural characteristics from the others. These 11 regions are marked with black border lines.} \label{fig:seoul-intro}
\end{figure}

The two datasets shown in Figure \ref{fig:seoul-intro} are the movements on weekdays and weekends in December 2023, respectively. Using these datasets, two directed graphs are created, with each vertex representing one of the 424 regions in Seoul. By applying the measures proposed in this study, we can readily identify popular and crowded regions and those that exhibit different structural behavior from the others. For example, in Figure \ref{fig:seoul-intro}(a), the five regions with the highest $L_1$ prestige are marked with black borders. These are well-known commercial districts in Seoul. Figure \ref{fig:seoul-intro}(b) shows the 11 northern regions where the difference between the highest and lowest local $L_1$ prestige values exceeds 0.6. These regions have unique relationships with the `hub' regions of Seoul. We validate this observation and give a comprehensive discussion in Section \ref{sec:seoul}, along with details on these datasets and graph construction. 

All proofs are deferred to the Appendix. Software for reproducing the analysis in this paper is available as an R package \textbf{L1centrality} \citep{l1cent} and at \url{https://github.com/seungwoo-stat/L1prestige-paper}. 

\section[L1 Prestige Measure]{$L_1$ Prestige Measure} \label{sec:l1pres}

Let $G = (V,E)$ denote a directed graph, where $V = \{v_1,\dots,v_n\}$ is a set of vertices, and $E$ is a set of directed edges. Denote $\eta_i$s as nonnegative weights of $v_i$s that represent the importance of each vertex. We shall call this vertex weights as \emph{multiplicities} and assume $\etadot\coloneqq\sum_{j=1}^n \eta_j >0$. Furthermore, we assign each edge a positive weight, which represents the length of that edge. Define $d(v_i, v_j)$, the \emph{geodesic distance} from $v_i$ to $v_j$, as the shortest path length. Here, the path length refers to the sum of the weights of the edges in that path. If there is no path from $v_i$ to $v_j$, $d(v_i,v_j)$ is set to infinity. In this paper, we restrict our attention to strongly connected graphs, i.e., graphs where every vertex can be reached from any other vertex. This restriction makes any geodesic distance from one node to another finite. A possible approach to alleviate this restriction is discussed in Section \ref{sec:discussion}. We note that $d(\cdot,\cdot)$ is not necessarily a distance function in a mathematical sense. Specifically, $d(v_i, v_j)$ and $d(v_j, v_i)$ might not be equal; symmetry is violated. When edge weights and vertex multiplicities are all set to one, a typical unweighted graph can be represented. 

For a strongly connected graph $G = (V, E)$, we first define the notion of \emph{graph prestige median}. This refers to a (nonempty) set of vertices that minimize the weighted sum of geodesic distances from the other vertices to it:
\begin{align}
m\left(G; \frac{\eta_1}{\etadot},\dots,\frac{\eta_n}{\etadot}\right) = \left\{v_i\in V: \sum_{j=1}^n \frac{\eta_j}{\etadot} d(v_j, v_i) \leq \sum_{j=1}^n \frac{\eta_j}{\etadot} d(v_j, v_k),~ k = 1,\dots, n\right\}. \label{eq:median-set}
\end{align}
For simplicity, we refer to the element of $m(G; \frac{\eta_1}{\etadot},\ldots,\frac{\eta_n}{\etadot})$ as \emph{prestige median vertex}. This is our view of the most prominent vertex in receiving choices: the vertex proximal from the other vertices is considered the most important. When all multiplicities are equal, the graph prestige median is equivalent to the vertices with the largest proximity prestige \cite[Ch.5.3]{wasserman_faust_1994}. The term and definition of prestige median vertex are analogous to the $L_1$ median for multivariate data \citep{small1990survey} and the graph median for undirected graphs \citep{hakimi1964optimum,kang-l1cent-paper}.

Given the definition of the prestige median vertex, we define the \emph{$L_1$ prestige} of vertex $v_k$ as
\begin{align}
\operatorname{Pres}(v_k) \coloneqq 1 - \calS(G)\inf\left\{w\geq 0: v_k\in m\left(G; \frac{\eta_1}{\etadot},\dots,\frac{\eta_k}{\etadot} + w,\dots, \frac{\eta_n}{\etadot} \right)\right\}, \label{eq:l1pres}
\end{align} 
where $\calS(G) \coloneqq \min_{i\neq j} \{d(v_i, v_j)/d(v_j,v_i)\}$. Roughly speaking, we rank the prominence of vertices according to the minimum amount of multiplicity that must be incremented for that vertex to become a prestige median vertex. The prestige median vertex always has the $L_1$ prestige of 1 (the maximum value) since the minimum amount of multiplicity to make it a prestige median vertex is 0. In this sense, the larger the $L_1$ prestige, the more important the vertex is in receiving choices, i.e., prestigious. 

Multiplication of the minimum multiplicity by $\calS(G)$ and subtraction from 1 in \eqref{eq:l1pres} are for normalization of the $L_1$ prestige measure; make it belong to the range $[0,1]$ so that comparison of the measure across several graphs is possible. Looking closely, it can be readily checked that \eqref{eq:l1pres} is equivalent to the following expression,
\begin{align}
\operatorname{Pres}(v_k) = 1 - \calS(G)\max_{j\neq k}\left\{\frac{\sum_{i=1}^n \eta_i\{d(v_i, v_k) - d(v_i, v_j)\}}{\etadot d(v_k, v_j)}\right\}^+, \label{eq:l1pres2}
\end{align}
where $\{\cdot\}^+ = \max\{0,\cdot\}$. Applying the triangle inequality ($d(v_i, v_k)-d(v_i,v_j)\leq d(v_j,v_k)$; be cautious on the order of indices) to the numerator of \eqref{eq:l1pres2}, $\operatorname{Pres}(v_k)\in [0,1]$. Note that $\calS(G) \in (0,1]$ and the smaller the $\calS(G)$ is, the more asymmetric the geodesic distances are in $G$. Hence, $\calS(G)$ can be interpreted as a constant that quantifies the symmetry of the graph's geodesic distances. 

In a matrix notation, denoting $\bD_{n\times n} = (d(v_i,v_j))_{ij}$, $\bfeta = (\eta_1,\dots,\eta_n)^\top$, and $\mathbf{1}_n = (1,\dots,1)^\top$,
\begin{align}
(\operatorname{Pres}(v_1),\dots,\operatorname{Pres}(v_n))^\top = \mathbf{1}_n - \frac{\calS(G)}{\etadot}\texttt{rowmax}\left\{\frac{\bD^\top\bfeta\mathbf{1}_n^\top - \mathbf{1}_n\bfeta^\top\bD}{\bD}\right\}^+. \label{eq:l1pres-mat}
\end{align}
In the expression above, division by a matrix is conducted element-wise, \texttt{rowmax} indicates the maximum element from each row of the matrix, and $\{\cdot\}^+$ is applied to each element of the vector. In conducting the \texttt{rowmax} operation, diagonal elements are ignored.


\begin{thm}[Properties of $L_1$ prestige]\label{thm:property}
The $L_1$ prestige measure has the following properties:
\begin{itemize}
\item[\emph{(P1)}] \emph{Scale invariance:} It is invariant to (positive) multiplicative transformations of vertex multiplicities and edge weights.
\item[\emph{(P2)}] \emph{Maximality:} It is maximized to 1 if and only if the given vertex is the prestige median vertex. Also, if $\eta_k/\etadot \geq 1/(1+\calS(G))$, then $\operatorname{Pres}(v_k) = 1$. If $\eta_k/\etadot > 1/(1+\calS(G))$, $v_k$ is the unique vertex with $\operatorname{Pres}(v_k) = 1$.
\item[\emph{(P3)}] \emph{Minimum value:} $\operatorname{Pres}(v_k) \geq \min\{(1+\calS(G))\eta_k/\etadot,1\}$. 
\item[\emph{(P4)}] \emph{Special case:} If $d(v_i, v_j) = d(v_j, v_i)$ for all $v_i,v_j\in V$, it reduces to the $L_1$ centrality for undirected graphs in \cite{kang-l1cent-paper}.
\end{itemize}
\end{thm}

\begin{remark}[Interpretation] \label{rmk:relev}
Properties (P2) and (P3) suggest that as the multiplicity of a vertex increases, it is more likely that the vertex has a high $L_1$ prestige. Hence, the vertex multiplicity can be interpreted as contextual information or prior knowledge of its prestige. Moreover, the larger the $L_1$ prestige, the less multiplicity is required for that vertex to \emph{replace} the current prestige median vertex. In other words, a vertex with high $L_1$ prestige can be regarded as a more \emph{relevant} vertex to the prestige median vertex in the sense of the ability to replace the prestige median vertex with a small amount of multiplicity. In the following section, this useful interpretation of the $L_1$ prestige will be used to define the concept of $L_1$ prestige-based neighborhood and the local version of the $L_1$ prestige. This is the essential aspect that distinguishes the proposed measure from the measures in the literature.
\end{remark}

\begin{remark}[$L_1$ centrality] \label{rmk:cent}
Similarly, we can define $L_1$ centrality, which quantifies the importance of a vertex in \emph{giving} choices to other vertices. The \emph{centrality median vertex} and $L_1$ centrality can be easily defined by switching the order of vertices in the distance functions in \eqref{eq:median-set}--\eqref{eq:l1pres2}. In \eqref{eq:l1pres-mat}, the distance matrices $\bD$ and $\bD^\top$ are transposed. The $L_1$ centrality measure also satisfies the properties in Theorem \ref{thm:property} and the interpretation of Remark \ref{rmk:relev} applies as well. The $L_1$ centrality and $L_1$ prestige for a directed graph can be used together to examine the prominence of each vertex from two different perspectives. This will be discussed in Section \ref{sec:seoul}.
\end{remark}

\begin{remark}[Undirected graphs] \label{rmk:undir}
When viewing undirected graphs as special cases of directed graphs with symmetric edges (and symmetric edge weights), (P4) implies that the $L_1$ prestige and $L_1$ centrality are canonical extensions of the $L_1$ centrality for undirected graphs proposed by \cite{kang-l1cent-paper}. Hence, these measures form a family of prominence measures for directed and undirected graphs. 
\end{remark}

\section[Multiscale Network Analysis via Local L1 Prestige]{Multiscale Network Analysis via Local $L_1$ Prestige} \label{sec:multiscale}

The $L_1$ prestige is a global measure considering all other vertices when determining prestige. As previously mentioned, different aspects of the network may become apparent when viewing it locally. In this section, we establish the concept of a $L_1$ prestige-based neighborhood and localize the $L_1$ prestige measure by confining the evaluation of the measure to the neighborhood of each vertex. In doing so, we present a framework for multiscale network analysis that can capture the prestige of each vertex at different scales of locality, providing more complete information of the network.


\subsection[L1 Prestige-Based Neighborhood]{$L_1$ Prestige-Based Neighborhood}

Given a directed graph, it may seem straightforward to define a neighborhood of a specific vertex as those vertices with a small geodesic distance from (or to) the vertex of interest. However, this notion of neighborhood ignores the multiplicity of vertices and the graph structure beyond the connecting path. The $L_1$ prestige-based neighborhood to be defined considers the entire structure of the graph (edges, weights, and multiplicities) when determining the neighborhood. The motivation for the definition of the $L_1$ prestige-based neighborhood is provided in Remark 1. Vertices with high $L_1$ prestige can be regarded as a neighborhood of the prestige median vertex because they require less multiplicity to replace the prestige median vertex. However, this is only valid for deriving the neighborhood of the prestige median vertex and cannot be applied to other vertices.

To solve this problem, we use Theorem \ref{thm:property} (P2). Suppose that we add multiplicity $\etadot/\calS(G)$ to $\eta_k$ and denote it as a \emph{modified graph w.r.t.\ $v_k$}. In the modified graph, $v_k$ becomes the prestige median vertex because the proportion of the new multiplicity of $v_k$ in the modified graph satisfies $(\eta_k + \etadot/\calS(G))/\{\etadot(1+1/\calS(G))\} \geq 1/(1+\calS(G))$. If $\eta_k >0$, the inequality is strict, implying that $v_k$ is the unique prestige median vertex in the modified graph. We can observe that in the modified graph, vertices with high $L_1$ prestige are either (i) prestigious in the original graph or (ii) have a small distance from the $v_k$ (which we denote as \emph{prestigious w.r.t.\ $v_k$}), since these vertices have a small numerator when deriving $L_1$ prestige using \eqref{eq:l1pres2}. 


Denoting the \emph{locality level} $\alpha\in [0,1]$, the \emph{order $\alpha$ $L_1$ prestige-based neighborhood} of $v_k$, say $\calN_\alpha(v_k)$, is defined as a set of vertices with $L_1$ prestige in the modified graph w.r.t.\ $v_k$ that are greater or equal to the $\alpha$th sample quantile of the $L_1$ prestiges in the modified graph. In light of Remark \ref{rmk:relev}, about $n\alpha$ vertices relevant to the vertex $v_k$ (including itself, because $v_k$ has $L_1$ prestige of 1 in the modified graph w.r.t.\ $v_k$) are selected. When $\alpha = 1$, $\calN_\alpha(v_k) = V$. 

Specifically, the $L_1$ prestige-based neighborhood strikes a balance between the prestigious vertex in the original graph and the prestigious vertex w.r.t.\ the vertex of interest.
For example, consider a number of vertices with the same geodesic distance from the vertex of interest. If the neighborhood each vertex is determined solely by the value of geodesic distances, then no vertex is more preferable than others as a neighborhood of the vertex of interest, regardless of its involvement in the network. On the other hand, $L_1$ prestige-based neighborhood will select a vertex with higher $L_1$ prestige as a neighborhood of the vertex of interest before a vertex with lower $L_1$ prestige. That is, $L_1$ prestige-based neighborhood takes into account the entire graph structure.

Readers knowledgeable in multivariate statistics will see this approach as bearing resemblance to the Mahalanobis distance. It considers all the data points and their relationships in the computation of the distance, instead of relying solely on the original Euclidean distance. The $L_1$ prestige-based neighborhood can be compared to points that have a small Mahalanobis distance from the central point in a multivariate dataset.

\subsection[Local L1 Prestige Measure]{Local $L_1$ Prestige Measure}

The $L_1$ prestige-based neighborhood of $v_k$ are vertices that are prestigious in the original graph or are prestigious w.r.t.\ $v_k$. The local $L_1$ prestige measures how prestigious $v_k$ is among these prestigious vertices by confining the evaluation of $L_1$ prestige. Specifically, the \emph{order $\alpha$ local $L_1$ prestige} is defined by modifying \eqref{eq:l1pres2} as follows,
\begin{align}
\operatorname{Pres}_\alpha(v_k) \coloneqq 1 - \calS(G)\max_{j: v_j\in\calN_\alpha(v_k)\setminus \{v_k\}}\left\{\frac{\sum_{i: v_i\in\calN_\alpha(v_k)} \eta_i \{d(v_i, v_k) - d(v_i, v_j)\}}{d(v_k,v_j)\sum_{i: v_i\in\calN_\alpha(v_k)}\eta_i}\right\}^+. \label{eq:local-pres}
\end{align}
Thus, in the local $L_1$ prestige, the locality level $\alpha$ has a clear interpretation: it controls the evaluation of the prominence of each vertex to approximately $100\alpha$\% relevant vertices. The $L_1$ prestige of \eqref{eq:l1pres2} is a special case of local $L_1$ prestige with $\alpha = 1$. We call this particular case the \emph{global $L_1$ prestige}, as $\calN_1(v_k) = V$ for any $v_k$. By examining the $L_1$ prestige at multiple locality levels, more information about the original graph data is retained, and more structural behaviors can be identified. We refer to this as \emph{multiscale network analysis} utilizing local $L_1$ prestige. 

Notice that the properties of the $L_1$ prestige in Theorem \ref{thm:property} can be easily modified and applied for the local $L_1$ prestige measure. Property (P1) holds for the local measure as well. In (P2) and (P3), replace $\etadot$ to $\sum_{i:v_i\in\calN_\alpha(v_k)}\eta_i$. For (P4), refer to Remark \ref{rmk:localnb} below. 

When defining the local measure, we modified \eqref{eq:l1pres2} and confined the computation. Another possible way to define a local measure is to compute the $L_1$ prestige on the subgraph of the original, where each subgraph consists of the $L_1$ prestige-based neighborhood of each vertex. However, this has two problems: (i) there is no guarantee that the induced subgraph is strongly connected, which makes the measure computation infeasible, and (ii) one needs to compute new geodesic distances for each subgraph. In contrast, in \eqref{eq:local-pres}, the geodesic distances of the original graph are reused; there is no need to compute the new geodesic distance. This makes computations of the local measures less burdensome, and there is no such concern for computational infeasibility. 



\begin{remark}[$L_1$ centrality-based neighborhood and local $L_1$ centrality]\label{rmk:localnb}
Following Remark \ref{rmk:cent}, $L_1$ centrality-based neighborhood and local $L_1$ centrality can be defined as well. The $L_1$ centrality-based neighborhood is defined in the same way as the $L_1$ prestige-based neighborhood. Modify the graph by adding multiplicity $\etadot/\calS(G)$ to the vertex of interest. This makes the vertex of interest the centrality median vertex. Select top $100\alpha$\% vertices with high $L_1$ centrality in the modified graph. This neighborhood balances the central vertices in the original graph and central vertices w.r.t.\ the vertex of interest, similar to the $L_1$ prestige-based neighborhood. The local $L_1$ centrality is defined by confining the computation of $L_1$ centrality to the neighborhood, as in \eqref{eq:local-pres}. Of course, the local $L_1$ centrality can also be utilized for multiscale network analysis. As in Remark \ref{rmk:undir}, the $L_1$ centrality-based neighborhood and $L_1$ prestige-based neighborhood coincide for undirected graphs and are identical to the one defined in \cite{kang-l1cent-paper}. Thus, these local measures are canonical extensions of the local measure in \cite{kang-l1cent-paper}.
\end{remark}

\begin{remark}[Data depth and $L_1$ prestige]
It is worth noting that the $L_1$ data depth by \cite{vardi2000multivariate} serves as the basis for the $L_1$ prestige and $L_1$ centrality. Data depth is a concept for multivariate data that extends univariate rank in a center-outward manner \citep{liu1999multivariate}. The similarity in concept between centrality and prestige, which measure the prominence of vertices, and data depth, which measures the centralness of data points, motivates adopting the ideas of data depth. Similarly, the local extension of the $L_1$ prestige and $L_1$ centrality incorporates the concept of local depth by \cite{paindaveine2013depth}, which determines the neighborhood of each point in a depth-based manner and constrains the evaluation of the data depth within the neighborhood. 
This suggests that there are numerous additional statistical topics to explore pertaining to network data through the lens of data depth, as we did in this paper. For example, it would be worthwhile to investigate whether the use of data depth for classification and clustering of multivariate data \citep[e.g.,][]{jornsten2004clustering,li2012dd} can be adapted for $L_1$ prominence measures and network data.
\end{remark}

\section{Analysis of Seoul Flow Network} \label{sec:seoul}

Seoul, the capital of South Korea, is a densely populated metropolis with about 10 million citizens. The city's dynamics are fueled by the movement of these individuals, making it naturally intriguing to observe their patterns of congregation and dispersal. This section aims to illustrate that the $L_1$ centrality, $L_1$ prestige, and their local extensions can be utilized to gain valuable insights from the movements inside Seoul. 

As of 2024, the city of Seoul comprises 25 primary administrative divisions, each named with the suffix `gu.' These divisions are further subdivided into 426 secondary administrative divisions, each named with the suffix `dong.' 

The city of Seoul publicly provides a monthly dataset, `Seoul Mobility Flow Data' (referred to as `Seoul flow data' hereafter). This dataset contains information about the number of people moving around in Seoul, categorized by various demographic criteria. The data provide estimates of the number of individuals moving from one dong to another inside Seoul, which are derived from mobile phone signals. This information is categorized by age group (in five-year intervals), time of arrival (in one-hour increments), gender, and day of the week (e.g., Monday). For example, from the data, it is possible to observe the number of men aged 20 to 24 who traveled from one specific dong to another between 9 a.m.\ and 10 a.m.\ on Monday in a given month. 

We construct directed networks utilizing the Seoul flow data from December 2023 and use the proposed measures. The process of constructing a directed graph is as follows. Each of the 424 dongs\footnote{As noted above, as of 2024, there are 426 dongs in Seoul; however, the Seoul flow data is based on the 424 dongs present at the time of development of this dataset. 
} represents a vertex. The multiplicity of each vertex is set to the number of movements within each dong. A directed edge is built between two vertices if at least one movement exists from the initial vertex to the other vertex. The weight of this edge is set to the reciprocal of the number of movements, meaning that the more moves there are, the smaller the path distance on this edge. Due to privacy concerns, only the count of moves above three is publicly available for the Seoul flow data. The counts of movements greater than 0 and less than or equal to 3 are marked as an asterisk (*) in the raw data. When constructing the networks, we imputed these values to 2. We construct two strongly connected and directed networks. One represents the morning movements occurring on weekdays in December 2023 (Figure \ref{fig:seoul-intro} (a)), and the other represents the weekend movements in December 2023 (Figure \ref{fig:seoul-intro} (b)).

\subsection{Weekday Morning: The Commuter and Commercial Town} \label{subsec:seoul-weekday}

First, we focus on movements occurring in the morning (8 a.m.--10 a.m.) on weekdays in December 2023. We set the age range from 25 to 64. The global $L_1$ centrality and global $L_1$ prestige are computed for each of the 424 dongs in this flow network. 

In this network, nodes with high $L_1$ centrality are expected to correspond to commuter towns with many residents rather than industrial towns. This is because high $L_1$ centrality indicates that a node plays a crucial role in giving choices to other nodes. During weekday morning hours, commuter towns send people to commercial (industrial) towns as they commute to work. Conversely, vertices with high $L_1$ prestige will be commercial towns that receive a large influx of people from other commuter towns during the morning hours.
\begin{figure}
\center
\includegraphics[width=.5\textwidth]{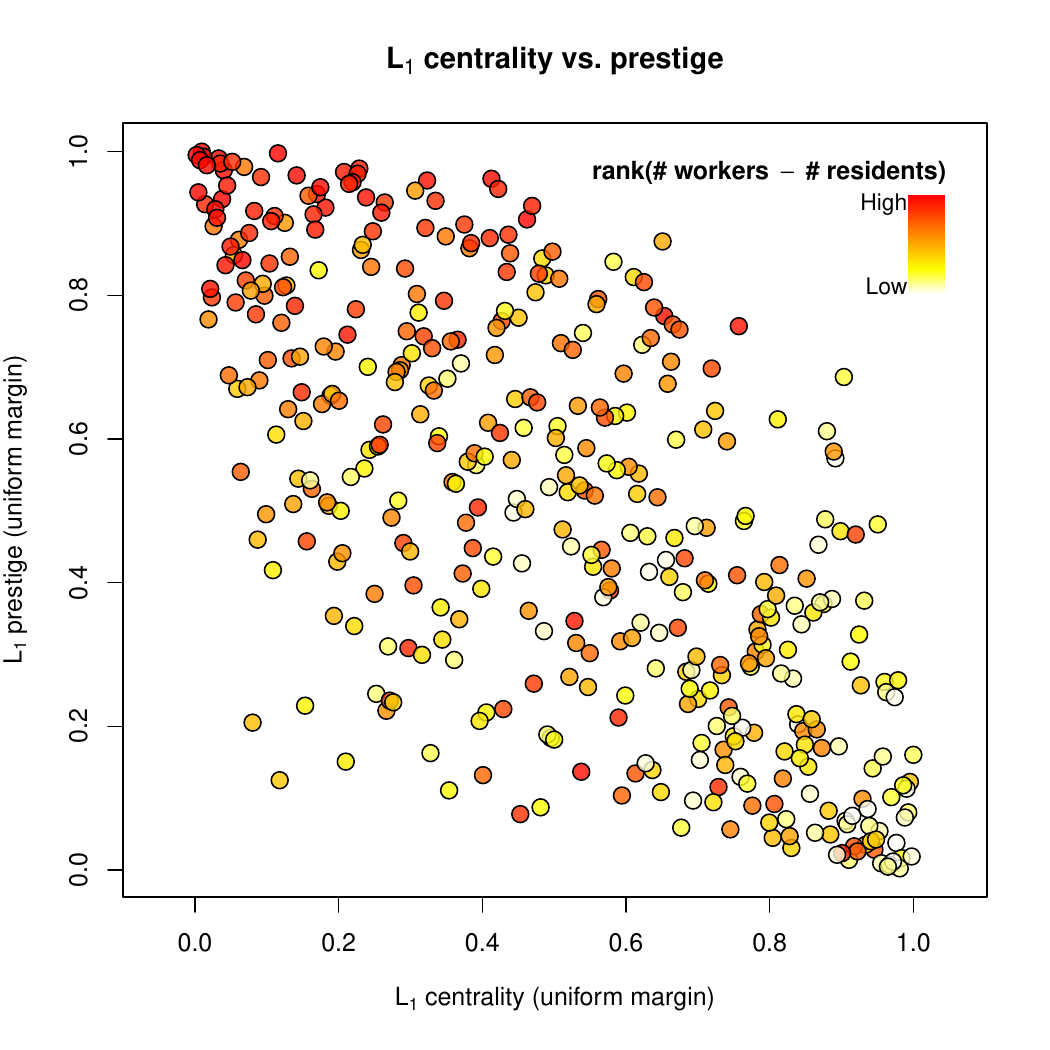}\vspace{-5mm}
\caption{Comparison of the $L_1$ centrality and $L_1$ prestige for 424 vertices in the weekday morning flow network of Seoul. Red points indicate areas with more employees and fewer residents and represent commercial towns. Yellow dots, on the other hand, indicate areas with more residents and fewer employees and represent commuter towns.} \label{fig:cent-vs-pres}
\end{figure}
In Figure \ref{fig:cent-vs-pres}, we can compare the $L_1$ centrality to the $L_1$ prestige for each vertex. For clear visualization, the $L_1$ centrality and prestige are transformed to have a uniform margin, i.e., the $i$th lowest $L_1$ prestige is transformed to $i/424$, and the same transformation is applied to the $L_1$ centrality. For this particular network, we observe from Figure  \ref{fig:cent-vs-pres} that vertices with a high $L_1$ centrality typically have a low $L_1$ prestige and \emph{vice versa}. The sample correlation coefficient of the two measures is $-0.70$ (with both measures transformed to have a uniform margin), indicating a strong negative relationship. The one-sided correlation test yields a $p$-value less than 0.001. 

We utilized two additional datasets for validation: (1) the number of employees at businesses in each dong as of 2021 and (2) the number of residents aged 25 to 64 in each dong as of the fourth quarter of 2023. The rank of the index `number of employees' minus `number of residents' at each vertex is represented by the colors in Figure \ref{fig:cent-vs-pres}. It is evident that vertices with a high $L_1$ prestige tend to rank high on this index, while vertices with a high $L_1$ centrality tend to rank low on this index. Specifically, the rank exhibits a negative sample correlation of $-0.53$ with the $L_1$ centrality (uniform margin) and a positive sample correlation of $0.58$ with the $L_1$ prestige (uniform margin). Both correlations are highly significant, with $p$-values less than 0.001 on the usual one-sided correlation test. In summary, the $L_1$ centrality and prestige measures effectively depict the behavior of commercial and commuter towns in this network.

\subsection{Weekend: The Far-North Land} \label{subsec:seoul-weekend}

The multiscale view of the Seoul flow network is more interesting: we examine the local $L_1$ prestige across various locality levels. Similar to the previous network, it focuses on analyzing the movements of individuals aged between 25 and 64 in December 2023. Yet, a network in this section is built for movements between 9 a.m. and 9 p.m. on the weekend. 

\begin{figure}
\center
\includegraphics[width = .6\textwidth]{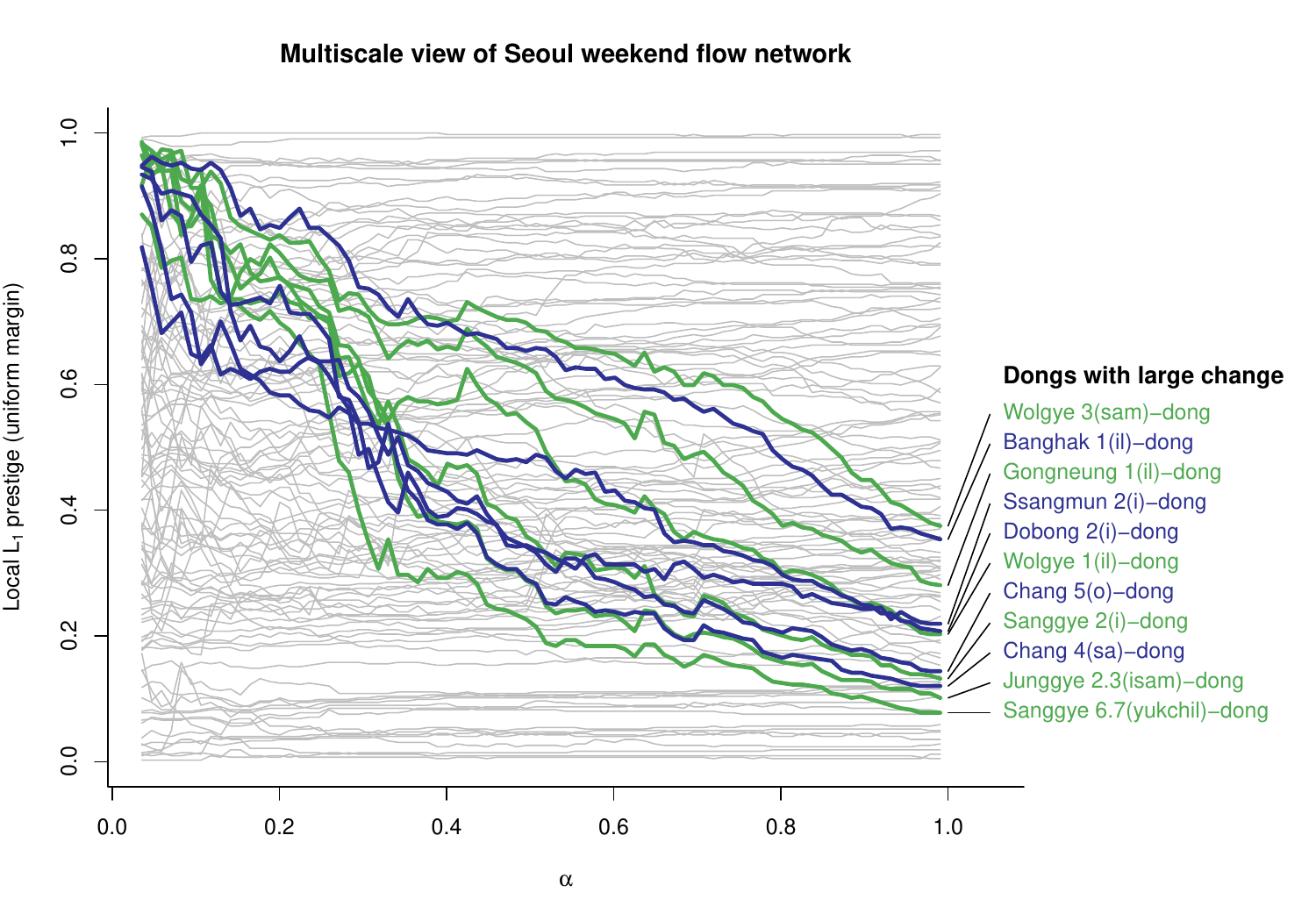}\vspace{-3mm}
\includegraphics[width = .39\textwidth, trim={15mm 0 0 15mm},clip]{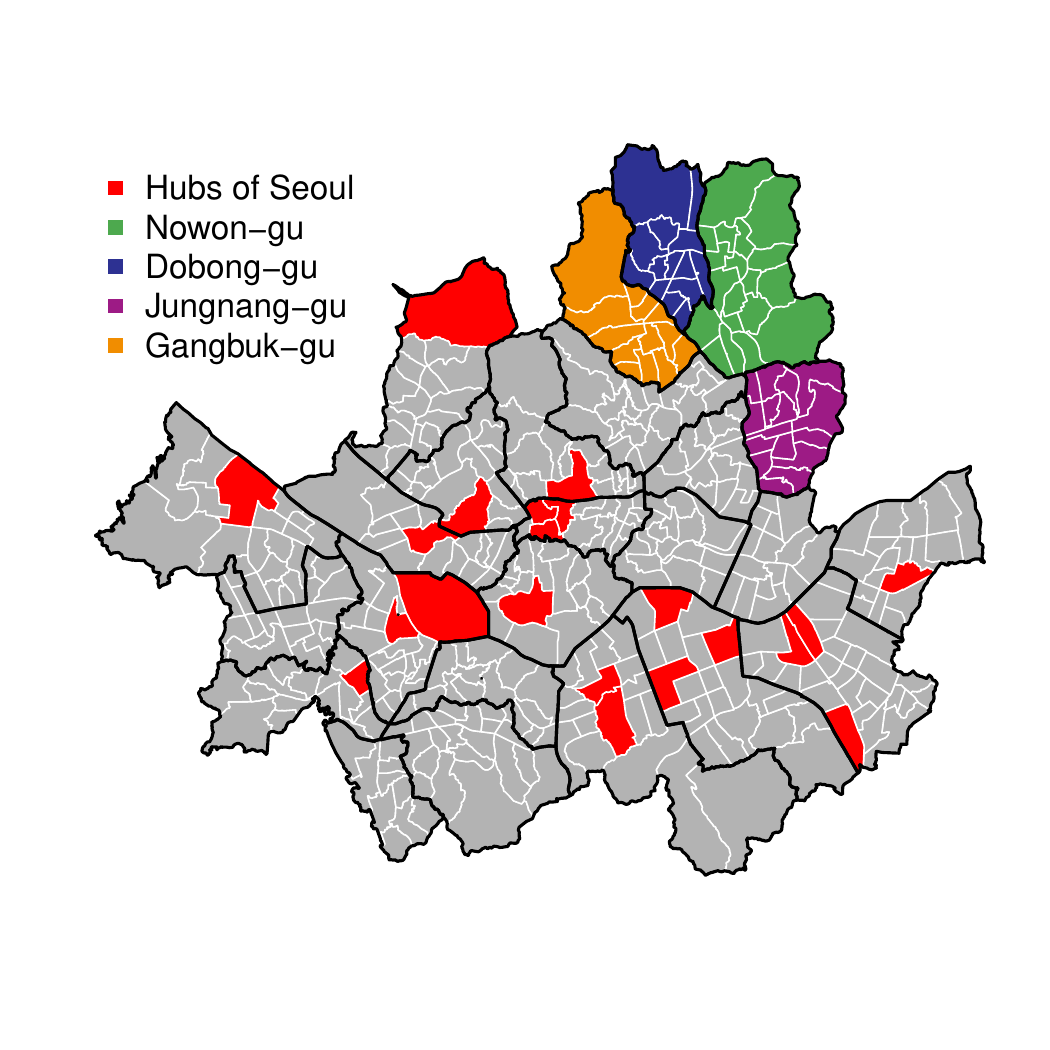}
\caption{(Left) Local $L_1$ prestige of each vertex (in uniform margin) at different levels of $\alpha$. Vertices with the highest and lowest local $L_1$ prestige value difference exceeding 0.6 are plotted with thick colored lines. Of these 11 vertices, the dongs in Nowon-gu and Dobong-gu are colored green and blue, respectively. We randomly selected 100 of the 413 vertices and represented them with thin gray lines for clear visualization. (Right) A map of Seoul divided into 424 dongs. The thick black lines are the boundaries for each gu. The dongs in four northern areas (Nowon-gu, Dobong-gu, Jungnang-gu, Gangbuk-gu) and 21 hubs are colored.}\label{fig:multiscale}
\end{figure}

We evaluated the local $L_1$ prestige for each of the 424 vertices at $\alpha = 15/424, 20/424, \dots, 420/424$. Note that global $L_1$ prestige is similar to the order 420/424 local $L_1$ prestige, so it is omitted. The local $L_1$ prestige was transformed to have a uniform margin for each level of $\alpha$. Each vertex in the Seoul weekend flow network is represented by a line in Figure \ref{fig:multiscale} (left panel). Vertices that show significant variation, especially those with a difference between their maximum and minimum local $L_1$ prestige values within the range greater than 0.6, are marked with thicker lines. Dongs corresponding to these vertices are also indicated by black border lines in Figure \ref{fig:seoul-intro} (b). Notably, all 11 vertices that show a significant change in local prestige values have a downward trend about $\alpha$. Furthermore, these vertices exist only in the two first-tier divisions: Nowon-gu (green lines) and Dobong-gu (blue lines). Put simply, certain dongs in Nowon-gu (6 out of 19) and Dobong-gu (5 out of 14) exhibit unique structural characteristics compared to others. 

We first counted the total number of incoming and outgoing movements for each dong. As a result, we found 22 outlying dongs with a large flow of incoming movements. We defined outliers as values that exceed 1.5 times the interquartile range (IQR) plus the third quartile or fall below the first quartile minus 1.5 times the IQR (lower fence). There were no outliers below the lower fence. Similarly, we identified 22 outliers with large outgoing movements, 21 of which overlap with the incoming movement outliers. These 21 vertices can be considered the \emph{hubs} of Seoul because they send and receive a lot of movements to and from other vertices. The hubs tend to have a high global $L_1$ prestige and a high global $L_1$ centrality. The 21 hubs are represented in red in Figure \ref{fig:multiscale} (right panel). 

\begin{table}
\caption{The maximum number of incoming (outgoing) movements from (to) any of the 21 hubs to (from) the dongs in each gu. The five divisions with the lowest counts are displayed, along with their ranking among the 25 first-tier divisions.}\label{tab:max-move}
\begin{center}
\begin{tabular}{r|rrrrr}
\hline
Division & Dobong-gu & Nowon-gu & Jungnang-gu & Gangbuk-gu & Gwanak-gu\\
\hline
Incoming count (rank) & 899.94 (1) & 1160.72 (2) & 1383.66 (3) & 1911.34 (4) & 3455.54 (5)\\
Outgoing count (rank) & 1308.14 (1) & 1591.82 (3) & 1365.35 (2) & 2246.50 (4) & 4219.78 (5)\\
\hline
\end{tabular}
\end{center}
\end{table}

Nowon-gu and Dobong-gu exhibit distinct behavior compared to other regions in relation to those 21 identified hubs: Nowon-gu and Dobong-gu have little interaction with the hubs, and the majority of the movements in and out of the two regions are confined to the northern region of Seoul, which does not possess any hubs.

Elaborating, Dobong-gu, Nowon-gu, Jungnang-gu, and Gangbuk-gu have the smallest maximum number of movements from (to) any of the 21 hubs to (from) the dongs located in each gu; this was about half of the fifth smallest value (Table \ref{tab:max-move}). These four divisions are geographically located in the northernmost part of Seoul, with Dobong-gu and Nowon-gu being the northernmost (Figure \ref{fig:multiscale}, right panel). The four divisions are distinctive from others in that they are the only first-tier divisions that do not possess any hubs and are not adjacent to divisions with hubs. Furthermore, Nowon-gu and Dobong-gu have more pronounced behaviors than the other two regions. The average proportion of incoming (outgoing) movements from (to) the four northernmost divisions to the total incoming (outgoing) movements is approximately 80\% (78\%) for Nowon-gu and Dobong-gu, while it is about 69\% (67\%) for Jungnang-gu and Gangbuk-gu. In other words, most of the flows from or to Nowon-gu and Dobong-gu primarily occur within the adjacent northern areas, regardless of the hubs.

This behavior explains the distinct multiscale curves shown in the left panel of Figure \ref{fig:multiscale}. Firstly, in the Seoul weekend flow network, the dongs in Nowon-gu and Dobong-gu will have long edges (large geodesic distances) from those not part of the four northern areas. This results in a small local $L_1$ prestige when $\alpha$ is close to one. In fact, of the 33 dongs in Nowon-gu and Dobong-gu, 29 have order 420/424 local $L_1$ prestige below the median. This includes the curves that are not colored in the left panel of Figure \ref{fig:multiscale}. 

Secondly, recall that the $L_1$ prestige-based neighborhood of each vertex is chosen by balancing between vertices that are prestigious in the original graph and vertices that are prestigious w.r.t.\ the vertex of interest. Since dongs in Nowon-gu and Dobong-gu do not have much movement to the hubs but send individuals to the four northern divisions a lot, vertices in Nowon-gu and Dobong-gu tend to choose vertices that correspond to dongs in the northern region as neighbors first, instead of choosing the globally prestigious hub vertices. Indeed, among the order $\alpha = 15/424$ $L_1$ prestige-based neighborhood of each vertex, the two divisions contained the fewest number of hubs on average: 1.11 for Nowon-gu and 1.36 for Dobong-gu, with the next smallest being 3.31. Moreover, on average, the order 15/424 $L_1$ prestige-based neighborhoods for dongs in Nowon-gu and Dobong-gu contain 7.47 and 10.93 dongs from the four northern divisions, respectively. In other words, the local $L_1$ prestige (with small $\alpha$) of the dongs in Nowon-gu or Dobong-gu is assessed mainly about other dongs in the northern divisions, resulting in a high local $L_1$ prestige of some vertices that are prominent among the dongs in the northern region. The order $15/424$ local $L_1$ prestige values (in a uniform margin) of the 33 dongs in Nowon-gu and Dobong-gu show a significant positive correlation with the total number of incoming movements from the four northern divisions (sample correlation coefficient 0.53, with $p$-value of the usual one-sided correlation test below 0.001), making it possible to interpret that colored vertices in Figure \ref{fig:multiscale} (left panel) are dongs that are popular (prestigious) in the northern region but have less interaction with the hubs.



\section{Discussion} \label{sec:discussion}

The $L_1$ prestige, centrality, and their local extensions, proposed in this study, have been demonstrated to be effective measures for assessing the prominence of vertices in strongly connected directed graphs. These measures can handle graphs with vertex and edge weights, and as a special case, they represent the $L_1$ centrality and its local version for undirected graphs previously developed by \cite{kang-l1cent-paper}. We have performed a multiscale network analysis that can provide a deeper understanding of network structures that may not be visible from a single-scale perspective. 

The tools developed in this paper have the potential to initiate further research in a variety of areas that use network data. For example, in the weekday flow network of the previous section, one might want to look at the shared characteristics of vertices with high $L_1$ centrality or $L_1$ prestige from a geographical or sociological perspective. Similarly, it would be interesting to investigate the underlying reason for the low interaction between the identified regions and the hubs in the weekend flow network as well as how urban planning may modify the current circumstance.

Further research on measures that apply to graphs that are not strongly connected would be interesting. We propose to define an alternate concept of graph median in \eqref{eq:median-set} that can be applied to a wider range of graphs, such as weakly connected or disconnected graphs. The prominence measure for a node can then be defined as the minimum multiplicity required to make it a median after appropriate normalization, as in Section \ref{sec:l1pres}. 







\section*{Disclosure Statement}

The authors declare no conflict of interest.

\section*{Data Availability Statement}


Data for reproducing the analysis in Section \ref{sec:seoul} are available at \url{https://github.com/seungwoo-stat/L1prestige-paper}. 

While all the data for replication are available in the GitHub repository mentioned above, the raw data can be downloaded via the Seoul Open Data Portal through the following links:
\begin{itemize}
\item Seoul Mobility Flow Data: \url{https://data.seoul.go.kr/dataVisual/seoul/seoulLivingMigration.do}
\item Statistics on the status of businesses in Seoul (number of employees): \url{https://data.seoul.go.kr/dataList/10598/S/2/datasetView.do} 
\item Statistics of the resident registration population in Seoul (number of residents): \url{https://data.seoul.go.kr/dataList/10727/S/2/datasetView.do}
\end{itemize}

\section*{Funding}

This research was supported by the National Research Foundation of Korea (NRF) funded by the Korea government (2021R1A2C1091357). 

\begin{appendix}
\section*{Appendix}
\section{Proof of Theorem \ref{thm:property}}
\begin{proof}
\begin{itemize}
\item[(P1)] It is immediate from \eqref{eq:l1pres2}.
\item[(P2)] In \eqref{eq:l1pres2}, observe that $\text{(numerator)}\leq (\etadot - \eta_k) d(v_j, v_k) -\eta_kd(v_k, v_j) = \etadot[d(v_j,v_k) - \eta_k/\etadot(d(v_j,v_k)+d(v_k,v_j))]$. If $\eta_k/\etadot\geq 1/(1+\calS(G))$, the numerator is non-positive because $d(v_k,v_j)\geq \calS(G)d(v_j,v_k)$. This yields that $\operatorname{Pres}(v_k) = 1$. 

For $k^\prime\neq k$, observe that
\begin{align*}
\operatorname{Pres}(v_{k^\prime}) &\leq 1 - \calS(G) \left\{\frac{\sum_{i=1}^n \eta_i\{d(v_i, v_{k^\prime}) - d(v_i, v_k)\}}{\etadot d(v_{k^\prime}, v_k)}\right\}^+\\
&\leq 1 - \calS(G) \left\{\frac{ - (\etadot - \eta_k) d(v_{k^\prime}, v_k) + \eta_k d(v_k, v_{k^\prime})}{\etadot d(v_{k^\prime}, v_k)}\right\}^+.
\end{align*}
It is immediately that the last expression is strictly smaller than 1 if $\eta_k/\etadot > 1/(1+\calS(G))$. Thus, $v_k$ is the unique vertex with the $L_1$ prestige 1.
\item[(P3)] Let $w = (1-(1+\calS(G))\eta_k/\etadot)/\calS(G)$. If $w\leq 0$, $v_k$ is the prestige median vertex by (P2) and $\operatorname{Pres}(v_k) = 1$. Suppose that $w > 0$. If the multiplicity of $v_k$ is incremented by $w$, i.e., from $\eta_k/\etadot$ to $\eta_k/\etadot + w$, $v_k$ becomes the graph prestige median due to (P2). Thus, \eqref{eq:l1pres} implies that $\operatorname{Pres}(v_k) \geq 1 - \calS(G)w = (1 + \calS(G))\eta_k/\etadot$.
\item[(P4)] If $d(v_i, v_j) = d(v_j, v_i)$ for all vertices, $\calS(G) = 1$ and \eqref{eq:l1pres} is equivalent to the definition of $L_1$ centrality of $v_k$ for an undirected graph in \cite{kang-l1cent-paper}, with the multiplicity and edge weight set to the same value as the given directed graph.  \qedhere
\end{itemize}
\end{proof}
\end{appendix}

\bibliographystyle{abbrvnat}
{\small\bibliography{bibs}}



\end{document}